# Shot Noise in Mesoscopic Systems with Resonance Andreev Tunneling

A. V. Lebedev and G. B. Lesovik

*Landau Institute for Theoretical Physics, Russian Academy of Sciences,
ul. Kosygina 2, Moscow, 117334 Russia*

The general scattering-matrix formalism is suggested for the description of shot noise in single-channel quantum NS contacts in the presence of smeared Andreev levels in the $eV < \Delta$ regime. It is shown that the noise spectral density as a function of frequency shows a number of characteristic features due to the presence of the Andreev resonances in the system. It turns out that, by analyzing these singularities, one can determine the relative phases of the Andreev reflection amplitudes corresponding to the neighboring resonances.

Shot noise in normal and superconducting systems is an interesting phenomenon, because it can be used to examine those features of electron transport which do not affect quantities of the type of current–voltage characteristic.

For example, the noise spectral density $S(\omega)$ of a normal quantum conductor has a singularity at frequency $\omega = eV/\hbar$ specified by the contact voltage [1, 2]. At the same time, shot noise of an ideal NS contact in the $eV < \Delta$ regime has a singularity at the Josephson frequency $\hbar\omega = 2eV$ [3, 4]. Since both these systems have well-defined linear current–voltage characteristics, it seems quite unexpected that the noise has a certain "intrinsic" frequency. This frequency is caused by a peculiar kind of two-particle interference, and it is precisely for this reason that it appears in the noise and not in conductance. In the $eV > \Delta$ regime, the noise shows additional features at frequencies $\hbar\omega = 2\Delta$ and $eV \pm \Delta$ (see [5, 6]) related to the singularity in the conductance at $eV = \Delta$.

The noise singularities at a finite frequency $\omega_0$ can also be revealed by studying the experimentally more accessible low-frequency noise in the presence of an additional ac voltage at frequencies $\Omega = \omega_0/n$ with integer $n$'s. For the normal contact and $eV = n\hbar\Omega$, singularities in the zero-frequency noise spectral density $S(0, V)$ were found in [2] in the presence of ac voltage at frequency $\Omega$ (this was confirmed experimentally in [7]). Similarly, $S(0, V)$ for an NS contact also shows singularities in the presence of ac voltage at frequency $\Omega$, but now $2eV = n\hbar\Omega$ [4], which was demonstrated experimentally in [8].

In [2, 4, 5], noise was studied using the scattering matrix formalism. However, this matrix was assumed to be independent of the particle energy.

In this work, we consider shot noise in superconducting systems in the Andreev regime $eV < \Delta$ at zero temperature, when the smeared Andreev energy levels (resonances) may be present, so that the scattering matrix of the system may *depend explicitly on energy*. It is found that the noise singularities can appear in this case at frequencies $(eV \pm \epsilon_n)/\hbar$ and $(\epsilon_n \pm \epsilon_m)/\hbar$.

It is remarkable (and unexpected) that the relative phases of the resonances can be judged from the presence of singularities at frequencies $(\epsilon_n \pm \epsilon_m)/\hbar$. Note that the problem of relative phases is not all that trivial. Experiments [9] demonstrate that the relative phase of the transmission amplitudes corresponding to two neighboring resonances in a quantum dot is a multiple of $2\pi$, instead of expected $\pi$ (a plausible explanation of this experiment is given in [10]). The revelation of the fact that the relative phases can be measured and the development of the general formalism for the description of noise in the presence of Andreev resonances are the major results of this work.

Moreover, in the particular case of the Andreev resonance near the Fermi level, a singularity appears at $\omega \simeq eV/\hbar$, which is typical of a normal conductor. This was likely observed experimentally in [11]. Our calculations show that the appearance of the frequency $\omega \simeq eV/\hbar$ in this case does not necessarily means that there is a transition to normal conductivity.

Both in the stationary case and in the presence of ac voltage, the calculation of noise using the scattering matrix technique is the well-known procedure and can be found in many works, so we will not go into the details of this calculation (see, e.g., [1, 2, 4, 5, 12] and references therein). We will assume that a single nor-mal metal contact is attached to the mesoscopic system under study. To calculate noise, one has to determine the scattering matrix for the contact. In the electron–hole space, this matrix has the form

$$\hat{\mathbf{S}}(\epsilon) = \begin{pmatrix} s_{ee}(\epsilon) & s_{eh}(\epsilon) \\ s_{he}(\epsilon) & s_{hh}(\epsilon) \end{pmatrix}. \quad (1)$$

We will not describe the internal organization of the mesoscopic system in detail but merely construct the scattering matrix using the unitarity and the electron–hole symmetry conditions (the latter follows from the Bogoliubov equation), assuming that the widths and positions of resonances are given by $E_n = \epsilon_n - i\Gamma_n/2$:

$$s_{ee}(\epsilon) = s_{hh}^*(-\epsilon), \quad s_{eh}(\epsilon) = -s_{he}^*(-\epsilon),$$
$$|s_{he}(\epsilon)| = |s_{eh}(\epsilon)|. \quad (2)$$

Since all resonances $E_n$ are simple poles of the $\hat{\mathbf{S}}$ matrix, the scattering matrix can be sought in the form

$$\hat{\mathbf{S}}(\epsilon) = \hat{\mathbf{S}}_0 + \sum_n \left( \frac{\hat{\mathbf{S}}_{-n}}{\epsilon - E_{-n}} + \frac{\hat{\mathbf{S}}_n}{\epsilon - E_n} \right), \quad (3)$$

where $\hat{\mathbf{S}}_0$ and $\hat{\mathbf{S}}_n$ are the constant $2 \times 2$ matrices, and $E_{\pm n} = \epsilon_n - i\Gamma_n/2$ are the resonances symmetric about zero.

Using symmetry conditions (2), one obtains the following relations between the matrix elements of $\hat{\mathbf{S}}_n$ and $\hat{\mathbf{S}}_0$:

$$\hat{\mathbf{S}}_0 = \begin{pmatrix} a_0 & b_0 \\ -b_0^* & a_0^* \end{pmatrix}, \quad \hat{\mathbf{S}}_n = \begin{pmatrix} a_n & b_n \\ b_{-n}^* & -a_{-n}^* \end{pmatrix}, \quad (4)$$

$$a_{-n} a_0^* = -a_n^* a_0; \quad b_{-n} b_0^* = -b_n^* b_0. \quad (5)$$

In turn, the unitarity condition gives $\hat{\mathbf{S}}_0^+ \hat{\mathbf{S}}_0 = \hat{\mathbf{1}}$ and

$$b_{-n}\left(a_0 + \sum_m \mathbf{A}_{nm} a_m\right) = a_{-n}\left(b_0 + \sum_m \mathbf{A}_{nm} b_m\right), \quad (6)$$

$$b_n^*\left(a_0 + \sum_m \mathbf{A}_{nm} a_m\right) = -a_n^*\left(b_0 + \sum_m \mathbf{A}_{nm} b_m\right), \quad (7)$$

where $\mathbf{A}_{nm} = 1/(E_n^* - E_m)$.

Let none of the numbers $a_n$ and $b_n$ be zero. By introducing the parameters $\alpha_n = b_n/a_n \neq 0$, one obtains from Eqs. (6) and (7) the following equations for the coefficients of the scattering matrix:

$$\alpha_n \alpha_{-n}^* = -1,$$
$$\sum_m \mathbf{B}_{nm} a_m + a_0 + b_0 \alpha_n^* = 0, \quad \forall n \quad (8)$$

where $\mathbf{B}_{nm} = \mathbf{A}_{nm}(1 + \alpha_n^* \alpha_m)$. One can rigorously show that, if $\forall \alpha_n \neq 0$, then the matrix $\mathbf{B}$ always has its inverse; i.e., Eq. (8) always has a solution for the $a_n$ numbers.

Using relationships (5), one can verify that the parameters $\alpha_n$ are not arbitrary but equal to

$$\alpha_{\pm n} = \pm \lambda_n e^{i\phi}, \quad (9)$$

where $\lambda_n = \pm 1$, and $e^{i\phi} = i\sqrt{(b_0/b_0^*)(a_0^*/a_0)}$ is a constant phase factor.

Equation (8) and condition (9) completely solve the problem of constructing the scattering matrix. Let $2N$ symmetric resonances $E_{\pm n} = \pm\epsilon_n - i\Gamma_n/2$ occur near the Fermi level. Then, by choosing $n$ numbers $\lambda_n = \pm 1$ for each of the $N$ resonances on one side of the Fermi level and specifying the unitary matrix $\hat{\mathbf{S}}_0$, one uniquely finds all coefficients of the matrices $\hat{\mathbf{S}}_n$ using Eq. (8) and conditions (4):

$$\hat{\mathbf{S}}_n = a_n \begin{pmatrix} 1 & \lambda_n e^{i\phi} \\ \lambda_n e^{-i\phi} & 1 \end{pmatrix}. \quad (10)$$

In the case of narrow resonances, i.e., if $|\epsilon_n - \epsilon_m| \ll \Gamma_n, \Gamma_m \, \forall n, m$, the parameterization using numbers $\lambda_n$ has a clear meaning: if $\lambda_n \lambda_{n+1} = \mp 1$ for two neighboring well-separated resonances $\epsilon_n$ and $\epsilon_{n+1}$, then the phase of the Andreev reflection amplitude $s_{he}$ or $s_{eh}$ changes, respectively, by $\pi$ or $2\pi$ upon going from one resonance to another.

Let us consider a noise at positive frequencies. One can obtain the following expression for the noise at $0 < \omega < 2eV/\hbar$ [$S(\omega) = 0$ at $\omega > 2eV/\hbar$] in terms of the Andreev and normal reflection amplitudes:

$$S(\omega) = \frac{2e^2}{h}$$
$$\times \int_{\hbar\omega - eV}^{eV} (R_N(\epsilon) R_A(\hbar\omega - \epsilon) + K(\epsilon) K^*(\hbar\omega - \epsilon)) d\epsilon, \quad (11)$$

where $R_A(\epsilon) = |s_{he}(\epsilon)|^2$, $R_N(\epsilon) = |s_{ee}(\epsilon)|^2$, and $K(\epsilon) = s_{ee}^*(\epsilon) s_{he}(\epsilon)$.

Using Eq. (10), one obtains the following expression for the noise:

$$S(\omega) = 2\frac{e^2}{h} \int_{\hbar\omega - eV}^{eV} d\epsilon$$

$$\times \left\{ R_A(\epsilon) - \left( \sum_{nm} \frac{\chi_n \chi_m (1 + \lambda_n \lambda_m)}{(\epsilon - E_n^*)(\hbar\omega - \epsilon - E_m^*)} + \text{c.c.} \right) \right. \quad (12)$$

$$\left. + \left( \sum_{nm} \frac{\chi_n^* \chi_m (1 - \lambda_n \lambda_m)}{(\epsilon - E_n)(\hbar\omega - \epsilon - E_m^*)} + \text{c.c.} \right) \right\},$$

where $\chi_n = a_n^* \sum_m A_{nm} \lambda_m a_m$ is a constant factor and $eV$ is the contact voltage. Although the integral in Eq. (12) can be elementarily evaluated, we will not present here the results, because the corresponding expressions are rather cumbersome and uninformative.

Certain general conclusions can be drawn for the narrow resonances. Indeed, for $|\epsilon_n - \epsilon_m| \ll \Gamma_n, \Gamma_m \; \forall n, m$, one can set $a_0 = 1$ and $b_0 = 0$ in the $\hat{S}_0$ matrix. Such a choice for $\hat{S}_0$ corresponds to the case where the Andreev reflection probability in the resonance is unity, as can easily be seen if one uses the well-known Breit–Wigner formula and expands $\hat{S}(\epsilon)$ near the respective resonance. Then $a_n = -i\Gamma_n/2$ and $\chi_n = -(i/4)\lambda_n \Gamma_n \; \forall n$. Taking also into account that the terms in the second sum of Eq. (12) are the products of two poles located on one side of the real axis and, hence, make a negligible contribution to the integral, one can obtain the following approximate formula for the noise at $\omega > 0$:

$$S(\omega) = 2\pi \frac{e^2}{h} \left\{ \sum_n \frac{\Gamma_n}{2} \right.$$

$$\left. - \frac{1}{8} \sum_{n,m} \frac{(1 + \lambda_n \lambda_m) \Gamma_n \Gamma_m (\Gamma_n + \Gamma_m)}{(\hbar\omega - \epsilon_n - \epsilon_m)^2 + (\Gamma_n + \Gamma_m)^2/4} \right\}, \quad (13)$$

where the summation is over only those resonances for which $\hbar\omega - eV < \epsilon_n < eV$.

It follows that the finite-frequency noise has singularities of two characteristic types: "steps" at frequencies $\hbar\omega_n = eV \pm \epsilon_n$ and "dips" with widths $\Gamma_n + \Gamma_m$ at frequencies $\omega_{nm} = (\epsilon_n + \epsilon_m)/\hbar > 0$ such that $\lambda_n \lambda_m = 1$, where the noise is partially suppressed (in some cases, almost to zero; see below). Inasmuch as $\lambda_n^2 = 1 \; \forall n$, the singularity at $\omega_{nn} > 0$ is always of the second type, whereas the appearance of such singularities at frequencies $\omega_{nm}$ depends on the relative phase of the Andreev reflection amplitudes for the $n$th and $m$th resonances.

If $\lambda_n \lambda_m = -1$ (relative phase of the Andreev reflection amplitudes for the corresponding resonances is a multiple of $\pi$; i.e., resonances are "in antiphase"), then the dip appears at frequency $\omega = |\epsilon_n - \epsilon_m|/\hbar$; in turn, if $\lambda_n \lambda_m = 1$ (relative phase is a multiple of $2\pi$; i.e., the resonances are "in phase"), then noise is suppressed at frequency $\omega_{nm} = (\epsilon_n + \epsilon_m)/\hbar$.

As for steps, the singularities of this type are always present, regardless of the relative phase of resonances. Although function (13) has discontinuities at $\omega_n = (eV + \epsilon_n)/\hbar$, the noise is continuous at these frequencies, because steps have finite widths of order $\Gamma_n$. Therefore, Eq. (13) properly describes the behavior of noise away from steps, i.e., at $eV + \epsilon_n < \hbar\omega < eV + \epsilon_{n+1}$.

One can show that the noise singularities at negative frequencies are exactly the same as at $\omega > 0$. Likewise, the positions of noise singularities of the second type (dips) are determined by the relative phases of Andreev resonances. For this reason, the symmetrized current correlator $S_{\text{sym}}(\omega) = 1/2(S(\omega) + S(-\omega))$ must show the same singularities and the same dependence of their positions on the relative phases of resonances. Therefore, noise measurements can provide information not only on the position of resonances but also on their relative phases, which cannot be done by measuring only the conductance of a mesoscopic system.

If all resonances have almost the same widths and are arranged periodically (within the widths $\Gamma_n$ and $\Gamma_{n+1}$ of the corresponding resonances), i.e., if $\epsilon_{n+1} - \epsilon_n \approx \Delta_\epsilon$, and if, moreover, any two neighboring resonances are in antiphase, then, as follows from Eq. (13), the noise at positive frequencies $\omega_n = n\Delta_\epsilon/\hbar$ is suppressed almost to zero because of the superposition of singularities at frequencies $\omega_{nn} = (\epsilon_n + \epsilon_n)/\hbar \approx (2n - 1)\Delta_\epsilon/\hbar$ and $\omega_{ij} = (\epsilon_i + \epsilon_j)/\hbar \approx \omega_{nn}$ such that $\lambda_i \lambda_j = 1$. This situation occurs, for example, in a ballistic NINS contact if the NS interface is close to ideal.

We illustrate this by a particular case where only two resonances, $E_1 = \epsilon_1 - i\Gamma_1/2$ and $E_2 = \epsilon_2 - i\Gamma_2/2$, fall within the $eV$ interval (if there are several resonances in the system, we assume that their energies $\epsilon_n \gg eV$, i.e., that they do not affect the noise behavior at frequencies $|\omega| \leq 2eV/\hbar$). All characteristic frequencies can conveniently be schematized by the energy diagram shown in Fig. 1. In this case, the scattering matrix can be parameterized only in two ways: $\lambda_1 = \lambda_2 = 1$ (in-phase resonances) and $\lambda_1 = -\lambda_2 = 1$ (antiphase resonances). The results of numerical calculation of the noise at positive frequencies are presented for each of the parameterizations in Fig. 2, where the singularities are clearly seen at all characteristic frequencies shown in Fig. 1. One can see from Fig. 1 that (cf. [11]), in the presence of a resonance near $\epsilon_F$, a step appears at frequency $\omega \simeq eV/\hbar$, which is not associated with any normal (nonsuperconducting) transport regime.

We now describe qualitatively how a finite temperature influences the noise singularities. At $T \neq 0$, the Fermi step is smeared; i.e., the characteristic energy $eV$ in the diagram in Fig. 1 is smeared by $\sim T$ (similar to the smearing of the resonances by $\sim \Gamma$). At $T \ll \min|\epsilon_n - \epsilon_m|$, the steps at frequencies $(eV \ll \epsilon_n)/\hbar$ are broadened to $\Gamma_n + T$ (Fig. 1), and the singularity at the Josephson frequency $2eV/\hbar$ acquires a width of $\sim 2T$. As for the singu-larities of the second type (dips) at frequencies $(\epsilon_n + \epsilon_m)/\hbar$, $\epsilon_n, \epsilon_m \ll eV$), their widths and shapes should not change at temperatures $T \ll \min|\epsilon_n - \epsilon_m|$.

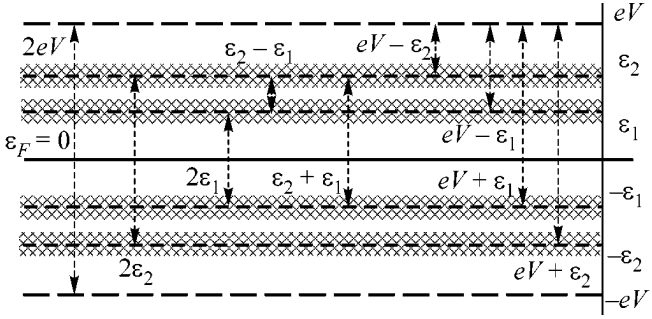

**Fig. 1.** Diagram of the characteristic energy intervals in the system. The corresponding frequencies are indicated by arrows, and the resonance widths are cross-hatched (scale is not kept).

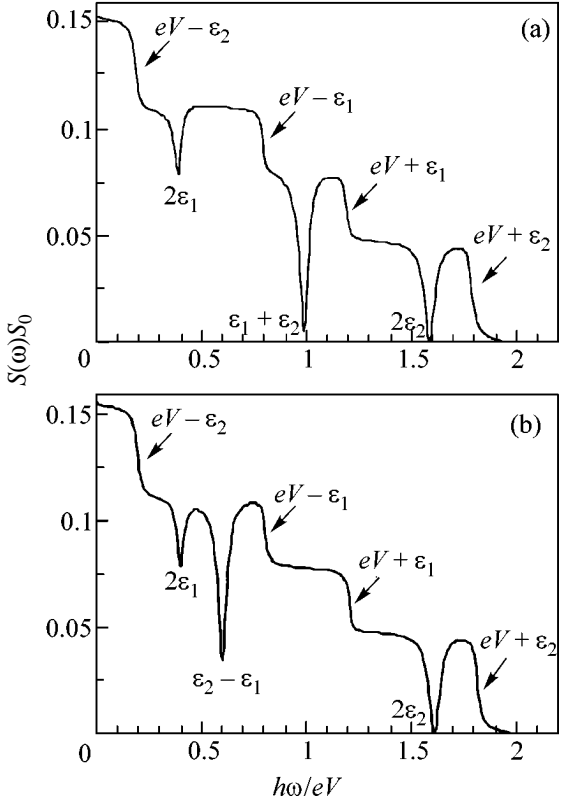

**Fig. 2.** Noise at positive frequencies in units of $S_0 = 2(e^2/h)eV$ for (a) in-phase levels and (b) antiphase levels: $\epsilon_1 = 0.2$ eV, $\epsilon_2 = 0.8$ eV, $\Gamma_1 = 0.02$ eV, and $\Gamma_2 = 0.03$ eV.

One can see that the noise has no zero-frequency singularities in the case of narrow resonances. Now let any two neighboring resonances, $\epsilon_1$ and $\epsilon_2$, be in antiphase and separated by a distance which cannot be regarded as small (i.e., $\epsilon_2 - \epsilon_1 \sim \Gamma_1, \Gamma_2$), but both them are away from the Fermi level ($\epsilon = 0$). In this case, one cannot use the Breit–Wigner formula for calculating the scattering amplitudes near $\epsilon_1$ and $\epsilon_2$. Whereas $R_A(\epsilon)$ for narrow resonances has maxima exactly at $\epsilon_n$, such is not the case for two closely spaced resonances. These resonances merge; i.e., the maxima of $R_A(\epsilon)$ start to approach each other and occur at the points $(1/2)(\epsilon_1 + \epsilon_2) \pm (1/2)\sqrt{(\epsilon_2 - \epsilon_1)^2 - \Gamma_1\Gamma_2}$ separated by a distance smaller than the distance between the corresponding quasi-discrete energy levels. The resonances fully merge at $\Delta = \epsilon_2 - \epsilon_1 = \sqrt{\Gamma_1\Gamma_2}$, resulting in a single maximum (equal unity) of $R_A(\epsilon)$ at $\epsilon = (\epsilon_1 + \epsilon_2)/2$. On further decrease in $\Delta$, this maximum decreases to zero. Therefore, after the resonances fully merge, measurement of conductance alone will not give an answer to the question as to whether two closely spaced quasi-discrete levels or only one level occur at $(\epsilon_1 + \epsilon_2)/2$.

One should expect that if the $\Delta$ distance becomes too small ($\Delta \sim \sqrt{\Gamma_1\Gamma_2}$), then the corresponding noise singularities at frequencies $2\epsilon_{1,2}/\hbar$, $(eV + 2\epsilon_{1,2})/\hbar$, etc., also merge. In particular, the singularity at $(\epsilon_2 - \epsilon_1)/\hbar$ shifts exactly to zero. The analytic expression for the low-frequency noise at $(\epsilon_1 + \epsilon_2)/2 \ll eV$ has the form

$$S(\omega) = 4\pi \frac{e^2}{h} \Gamma_+ \frac{\Delta^2 + \Gamma_-^2}{\Delta^2 + \Gamma_+^2}$$

$$\times \left\{ 1 - \frac{\Gamma_1\Gamma_2}{2\Gamma_+} \left( \frac{\Gamma_+ \cos 2\phi + (\hbar\omega + \Delta)\sin 2\phi}{(\hbar\omega + \Delta)^2 + \Gamma_+^2} \right. \right. \quad (14)$$

$$\left. \left. + \frac{\Gamma_+ \cos 2\phi + (\Delta - \hbar\omega)\sin 2\phi}{(\hbar\omega - \Delta)^2 + \Gamma_+^2} \right) \right\},$$

where $\Gamma_+ = (\Gamma_1 + \Gamma_2)/2$, $\Gamma_- = (\Gamma_1 - \Gamma_2)/2$, and $\phi = \arctan(\Gamma_+/\Delta)$. The results of the numerical calculation of the noise near zero frequency are presented in Fig. 3 as functions of the ratio $\Gamma_+/\Delta$. One can see that, at $\Gamma_+/\Delta < 1$ ("near-narrow" resonances), noise has a minimum near (but not at) zero; at $\Gamma_+/\Delta \sim 1$ the minimum shifts to zero; and at $\Gamma_+/\Delta \gg 1$ the noise has a maximum at zero.

Let us now briefly consider the nonstationary case with an alternating vector potential $A_x(t) = A_x \sin\Omega t$ applied to a restricted region of width $d$ near the contact. We are interested in the dependence of the zero-frequency noise $S_0(\Omega, V)$ on the frequency of ac potential and the contact voltage. Assuming that electrons passing through the region with ac potential gain an additional phase $\Phi(t) = \Phi \sin\Omega t$ ($\Phi = 2\pi \int dx A_x/\Phi_0$) and that the potential $A_x(t)$ weakly changes in a time of electron passage through the scattering region, one can obtain

$$S_0(\Omega, V) = \sum_{n=-\infty}^{\infty}\left(1 - 2\Theta\left(n - \frac{2eV}{\hbar\Omega}\right)\right) J_n^2(2\Phi) F(n\Omega), \quad (15)$$

where $J_n(x)$ is the Bessel function and $F(\omega) = 2e^2/h \int_{\hbar\omega - eV}^{eV}(R_A(\hbar\omega - \epsilon)R_N(\epsilon) + K(\epsilon)K^*(\hbar\omega - \epsilon))\,d\epsilon$ [see Eq. (11)]. Since $F(2eV) = 0$, the first derivative of the noise with respect to the voltage is equal to

$$\frac{\partial S_0(V, \Omega)}{\partial V} = \sum_{n=-\infty}^{\infty}\left(1 - 2\Theta\left(n - \frac{2eV}{\hbar\Omega}\right)\right) \times J_n^2(2\Phi)\frac{\partial F(n\Omega)}{\partial V}. \quad (16)$$

It is seen that the derivative $\partial S_0/\partial V$ as a function of $V$ has discontinuities (steps) at $2eV = n\Omega$. This result was obtained in [2] for the nonresonant case. In the resonance case, additional features appear as small peaks at $eV = \epsilon_i$ and $eV = n\Omega \pm \epsilon_i$, with widths $\Gamma_i$ and heights given by the Bessel function squared $J_n^2(2\Phi)$. It should be noted that both the step heights at $2eV = n\hbar\Omega$ and the peak heights at $eV = \epsilon_i$ and $n\hbar\Omega \pm \epsilon_i$ are sensitive to the phase incursion during the electron (hole) passage through the region with the ac potential.

This calculation is formally valid only if $\hbar\Omega \ll \Gamma$ (the ac potential weakly changes during the scattering time). Because of this, the results obtained are formally invalid at $\hbar\Omega > \Gamma$. However, we believe that, at a qualitative level, they correctly describe the behavior of $\partial S(0)/\partial V$ in this case as well (at least, this relates to the positions of the peak and step singularities).

We thank the participants in the seminar of the Quantum Mesoscopics Division of the Institute for Theoretical Physics RAS for discussions. This work was supported by the Russian Foundation for Basic Research (project no. 00-02-16617), the Ministry of Science (project "Physical Principles of Quantum Calculations"), the Dutch Scientific Foundation (grant for collaboration with Russia), and the Swiss Scientific Foundation.

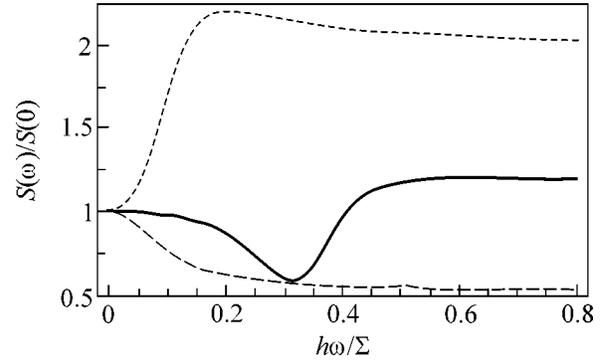

**Fig. 3.** Noise near $\omega = 0$ for the case of closely spaced levels $\Gamma_1 = \Gamma_2 = 0.1\Sigma$, where $\Sigma = (\epsilon_1 + \epsilon_2)/2$: $\Gamma_+/\Delta =$ (solid line) 0.25, (finely dashed line) 1, and (coarse-dashed line) 10.